\documentclass[12pt]{article}
\usepackage[dvips]{color}
\usepackage{epsfig}
\usepackage{amsmath}
\usepackage{graphicx}

\begin{document}
\title{\bf Langevin Diffusion Coefficients Ratio in STU Model with Higher Derivative Corrections}
\author{{B. Pourhassan$^{a}$\thanks{Email:
bpourhassan@yahoo.com}, M. Karimi$^{b}$\thanks{karimi.m290@yahoo.com}, S. Mojarrad$^{b}$\thanks{s.mojarad.sm@gmail.com}}\\
$^{a}${\small {\em  School of Physics, Damghan University, Damghan, 3671641167,  Iran.}}\\
$^{b}${\small {\em  Physics Department, Shahrood University of Technology, Shahrood,
Iran}}} \maketitle
\begin{abstract}
\noindent In this letter, we study Langevin diffusion coefficients for the five dimensional $\mathcal{N}=2$ STU model in presence of higher derivative corrections. We obtained effect of black hole charge, corresponding to the chemical potential, on the Langevin diffusion coefficients ratio. We confirm universal behavior of transverse to longitudinal ratio of coefficients.
{\bf Pacs:} 04.65.+e, 11.25.Mj, 12.38.Mh.\\\\
\end{abstract}

Study of quark-gluon plasma (QGP) using the AdS/CFT correspondence \cite{1, 2} is important subject in the last decade \cite{3}. According
to AdS/CFT correspondence there is a relation between a conformal field
theory (CFT) in $d$-dimensional space and a supergravity theory in
$(d+1)$-dimensional anti-de Sitter (AdS) space. It is indeed a relation between type IIB string theory in
$AdS_{5}\times S^{5}$ space and $\mathcal{N}=4$ super Yang-Mills theory on the 4-dimensional boundary of $AdS_{5}$ space. The study of the QGP is a testing ground
for finite temperature field theory, which are important to
understand the early evolution of the universe. The most important
quantities of QGP are drag force \cite{4, 5} and jet-quenching parameter \cite{6}.\\
The jet-quenching parameter obtained using the expectation value of a closed
light-like Wilson loop in the dipole approximation \cite{7}. In
order to calculate the jet-quenching
parameter in QCD one needs to use perturbation theory. But, by using the AdS/CFT correspondence, it can be calculated in non-perturbative quantum field
theory \cite{8}. On the other hand,  the drag force is indeed energy loss of moving heavy quark through
QGP. The motion of a heavy quark in context of QCD has dual picture in
the string theory, one can imagine an open string attached to the D-brane and
stretched to the black hole horizon. The stochastic motion related to the fluctuation correlations of the trailing string \cite{9,10}, which can be obtained in terms of the Langevin coefficients \cite{11, 12}. Therefore, the Langevin coefficients are important to study QGP. In the Ref. \cite{13} it has been found that the longitudinal Langevin diffusion coefficient along the string motion is larger than the transverse coefficient. Also, in the Ref. [14] the Langevin diffusion of a relativistic heavy quark in a general anisotropic
strongly coupled background has been studied.\\
In this letter, we would like to study Langevin diffusion coefficients in the STU model. The STU model includes a chemical potential to the model. For example, presence of a baryon number chemical potential for heavy quark in the context of AdS/CFT correspondence yields to introducing
a macroscopic density of heavy quark baryons. The STU model is
a kind of $D=5$, $\mathcal{N}=2$ gauged supergravity theory
which is dual to the $\mathcal{N}=4$ SYM theory with finite chemical
potential. The solutions of ${\mathcal{N}}=2$ supergravity may be
solutions of supergravity theory with more supersymmetry. It has been found that $\mathcal{N}=2$ the supergravity is an ideal
laboratory \cite{15}. Therefore, it may be to consider the STU model
as a gravity dual of a strongly coupled plasma. Moreover, the
$D=5$, $\mathcal{N}=2$ gauged supergravity theory is a natural way
to explore gauge/gravity duality, and three-charge non-extremal
black holes are important thermal background for this
correspondence. The STU model describes a five-dimensional
space-time which its four-dimensional boundary includes QCD. Drag force and jet-quenching parameter already obtained using the AdS/CFT in STU background and such studies called STU-QCD correspondence \cite{16,17,18,19,20}. Shear viscosity to entropy ratio of dual QGP also investigated in STU model \cite{Amani}. We will also study universal longitudinal and transverse Langevin coefficients ratio in presence of higher derivative corrections. STU model already used to study holographic superfluids and superconductors \cite{S1,S2}.\\
The STU model has generally 8-charged non-extremal black hole.
However, there are many situations of the black holes with four and three charge. In that case there is
great difference between the three-charged and four-charged black
holes. For example if there are only 3 charges, then the entropy
vanishes (except in the non-BPS case). So, one really needs four
charges to get a regular black hole. In 5 dimensions the situation
is different, there is no distinction
between BPS and non-BPS branch. So, in 5 dimensions the
three-charged configurations are the most interesting ones \cite{21}. Hence, our interest is the three-charged non-extremal black hole
solution in ${\mathcal{N}}=2$ gauged supergravity which is called
STU model described by the following solution \cite{22},
\begin{equation}\label{s1}
ds^{2}=-\frac{f_{k}}{{\mathcal{H}}^{\frac{2}{3}}}dt^{2}
+{\mathcal{H}}^{\frac{1}{3}}(\frac{dr^{2}}{f_{k}}+\frac{r^{2}}{R^{2}}d\Omega_{3,k}^{2}),
\end{equation}
where,
\begin{eqnarray}\label{s2}
f_{k}&=&k-\frac{\mu}{r^{2}}+\frac{r^{2}}{R^{2}}{\mathcal{H}},\nonumber\\
{\mathcal{H}}&=&\prod_{i=1}^{3} H_{i},\nonumber\\
H_{i}&=&1+\frac{q_{i}}{r^{2}}, \hspace{10mm} i=1, 2, 3,
\end{eqnarray}
where $R$ is the constant AdS radius and relates to the coupling
constant via $R=1/g$, and $r$ is the radial
coordinate along the black hole, so the boundary of AdS space
located at $r\rightarrow\infty$ (or $r=r_{m}$ on the D-brane). The
black hole horizon specified by $r=r_{h}$ which is obtained from
$f_{k}=0$. In the STU model there are three real scalar fields,
given by,
\begin{equation}\label{s3}
X^{i}=\frac{{\mathcal{H}}^{\frac{1}{3}}}{H_{i}},
\end{equation}
which is also solution of the metric (\ref{s1}) and satisfy the
following condition, $\prod_{i=1}^{3}X^{i}=1$. In another word, if
one set $X^{1}=S$, $X^{2}=T$, and $X^{3}=U$, then there is the special condition written as $STU=1$.  Finally, the factor of $k$
indicates the space curvature. The special case of $k=0$ corresponds to the black brane limit relevant to the thermal CFT in an infinite volume.\\
The Brownian motion of moving quark at fixed velocity understood using generalized Langevin coefficients. There are longitudinal and transverse Langevin coefficients which can be written in terms of world-sheet metric temperature given by \cite{13, 14},
\begin{equation}\label{4}
T=\frac{1}{4\pi}\sqrt{\frac{1}{G_{00}G_{rr}}(G_{00}G_{pp})^{\prime}(\frac{G_{00}}{G_{pp}})^{\prime}}|_{r=r_{0}},
\end{equation}
where prime denote derivative with respect to $r$ and,
\begin{eqnarray}\label{5}
G_{00}&=&-\frac{f_k}{H^{\frac{2}{3}}}=-\frac{k+\frac{(1+\frac{q_1}{r^2})(1+\frac{q_2}{r^2})(1+\frac{q_3}{r^2})r^2}{R^2}-\frac{\mu}{r^2}}{((1+\frac{q_1}{r^2})(1+\frac{q_2}{r^2})(1+\frac{q_3}{r^2}))^{\frac{2}{3}}}\\
G_{rr}&=&\frac{H^{\frac{1}{3}}}{f_k}=\frac{((1+\frac{q_1}{r^2})(1+\frac{q_2}{r^2})(1+\frac{q_3}{r^2}))^{\frac{1}{3}}}{k+\frac{(1+\frac{q_1}{r^2})(1+\frac{q_2}{r^2})(1+\frac{q_3}{r^2})r^2}{R^2}-\frac{\mu}{r^2}}\\
G_{kk}&=&\frac{H^{\frac{1}{3}}r^2}{R^2}=\frac{((1+\frac{q_1}{r^2})(1+\frac{q_2}{r^2})(1+\frac{q_3}{r^2}))^{\frac{1}{3}}r^2}{R^2}\\
G_{pp}&=&\frac{H^{\frac{1}{3}}r^2}{R^2}=\frac{((1+\frac{q_1}{r^2})(1+\frac{q_2}{r^2})(1+\frac{q_3}{r^2}))^{\frac{1}{3}}r^2}{R^2},
\end{eqnarray}
also $r_{0}$ is root of the following equation (obtained from reality condition),
\begin{equation}\label{7}
G_{00}+G_{pp}v^{2}=0
\end{equation}
which reduced to the following equation for the case of $k=0$,
\begin{eqnarray}\label{8}
0&=&(1-v^{2})r^{6}+\left(R^{2}+(q_{1}+q_{2}+q_{3})(1-v^{2})\right)r^{4}\nonumber\\
&+&\left((q_{1}q_{2}+q_{1}q_{3}+q_{2}q_{3})(1-v^{2})-\mu R^{2}\right)r^{2}+q_{1}q_{2}q_{3}(1-v^{2}).
\end{eqnarray}
It is clear that at $v=0$ limit $r_{0}=r_{h}$ where $r_{h}$ is black hole horizon radius given by $f_{k}=0$.
It has been argued that the longitudinal and transverse Langevin coefficients ratio written as follow,
\begin{eqnarray}\label{9}
\frac{\kappa_{\parallel}}{\kappa_{\perp}}&=&\pm{16\pi^{2}(\frac{G_{00}G_{rr}}{G_{kk}G_{pp}(\frac{G_{00}}{G_{pp}})^{\prime}\lvert(\frac{G_{00}}{G_{pp}})^{\prime}\lvert}})|_{r=r_{0}}T^2 \nonumber\\
&=&\frac{(G_{00}G_{pp})^{\prime}}{G_{kk}G_{pp}(\frac{G_{00}}{G_{pp}})^{\prime}}|_{r=r_{0}}.
\end{eqnarray}
After some calculation one can obtain,
\begin{equation}\label{10}
\frac{\kappa_{\parallel}}{\kappa_{\perp}}=\frac{r_{0}^{10}+Ar_{0}^{8}+Br_{0}^{6}+Cr_{0}^{4}+Dr_{0}^{2}+E}{ar_{0}^{8}+br_{0}^{6}+cr_{0}^{4}+dr_{0}^{2}+e},
\end{equation}
where we defined following coefficients,
\begin{eqnarray}\label{10-1}
A&=&\frac{5}{3}(q_{1}+q_{2}+q_{3})+\frac{1}{2}kR^{2},\nonumber\\
B&=&\frac{2}{3}(q_{1}^{2}+q_{2}^{2}+q_{3}^{2})+\frac{8}{3}(q_{1}q_{2}+q_{1}q_{3}+q_{2}q_{3})+\frac{2}{3}kR^{2}(q_{1}+q_{2}+q_{3}),\nonumber\\
C&=&(q_{1}+q_{2})q_{3}^{2}+(q_{1}+q_{3})q_{2}^{2}+(q_{2}+q_{3})q_{1}^{2}+\frac{5}{6}kR^{2}(q_{1}q_{2}+q_{1}q_{3}+q_{2}q_{3})\nonumber\\
&+&4q_{1}q_{2}q_{3}-\frac{1}{6}\mu R^{2}(q_{1}+q_{2}+q_{3}),\nonumber\\
D&=&\frac{4}{3}q_{1}q_{2}q_{3}(q_{1}+q_{2}+q_{3})+\frac{1}{3}(q_{1}^{2}q_{2}^{2}+q_{1}^{2}q_{3}^{2}+q_{2}^{2}q_{3}^{2})-\frac{1}{3}\mu R^{2}(q_{1}q_{2}+q_{1}q_{3}+q_{2}q_{3})+kR^{2}q_{1}q_{2}q_{3},\nonumber\\
E&=&-\frac{1}{2}q_{1}q_{2}q_{3}(\mu R^{2}-\frac{2}{3}(q_{1}q_{2}+q_{1}q_{3}+q_{2}q_{3})),\nonumber\\
a&=&-\frac{1}{2}kR^{2},\nonumber\\
b&=&\mu R^{2},\nonumber\\
c&=&\frac{1}{2}\left(\mu(q_{1}+q_{2}+q_{3})+k(q_{1}q_{2}+q_{1}q_{3}+q_{2}q_{3})\right)R^{2},\nonumber\\
d&=&kR^{2}q_{1}q_{2}q_{3},\nonumber\\
e&=&-\frac{1}{2}\mu q_{1}q_{2}q_{3}.
\end{eqnarray}

Already it is concluded that $\kappa_{\parallel}\geq\kappa_{\perp}$ is universal for isotropic backgrounds, where equality hold for $v=0$. It is also verified for the STU model with positive charge and illustrate for some situations summarized in the tables 1, 2 and 3. Without loss of generality, we set $R=\mu=1$ and vary black hole charges for three different cases of $k=0$, $k=1$ and $k=-1$. From the table 1 the case of $k=1$, table 2 for the case of $k=0$ and table 3 for the case of $k=-1$ we see universal behavior of $\frac{\kappa_{\parallel}}{\kappa_{\perp}}$ as expected. Here, we consider some unphysical cases of negative charge and take into account all possible mathematical situations, however universality hold for all cases.

\begin{center}
  \begin{tabular}{|@{} l @{} ||@{} c @{} | @{}r @{}|@{} c @{} |@{} c @{} |@{} c @{} ||@{} l @{} ||@{} c @{} | @{}r @{}|@{} c @{} |@{} c @{} |@{} c @{} |}
    \hline
    L           & $r_{0}$      & $q_{1}$  & $q_{2}$ & $q_{3}$ &   $v$     & L           &  $r_{0}$        & $q_{1}$ & $q_{2}$ & $q_{3}$  &   $v$ \\ \hline\hline
    1.0000006   & 0.786        & 0        &    0    &    0    &   0.001   & 1.67        &  0.058          &  1      &   1     &    0     &    0.1 \\ \hline
    3           &    1         & 0        &    0    &    0    &   1       & 3.54        &  0.74           &  1      &   1     &    0     &    0.9 \\ \hline
    1.0000008   & 0.64         & 1        &    0    &    0    &   0.001   &   $\infty$  &      1          & -1      &  -1     &    0     &    0.1 \\ \hline
    2.6         & 0.866        & 1        &    0    &    0    &   0.9     &   $\infty$  &      1          & -1      &   0     &    0     &    0.99 \\ \hline
    4.3         &    1         & 1        &    0    &    0    &   1       &   $\infty$  &      1          &  1      &   1     &   -1     &    0.99 \\ \hline
    4.7         & 0.58         & 10       &    0    &    0    &   0.9     & 1.0086      &  0.78           &  1      &  -1     &   -1     &    0.1 \\  \hline
    1.01        & 0.30         & 10       &    0    &    0    &   0.1     & 1.69        &  0.14           &  1      &  -1     &   -1     &    0.99 \\  \hline
     16.333     &       1      & -1       &   -1    &    0    &   0.9     & 1.000086    &  0.786          &  1      &  -1     &   -1     &    0.01 \\
    \hline
  \end{tabular}
\end{center}
Table 1. Value of $L=\frac{\kappa_{\parallel}}{\kappa_{\perp}}$ for $k=1$.

\begin{center}
  \begin{tabular}{|@{} l @{} ||@{} c @{} | @{}r @{}|@{} c @{} |@{} c @{} |@{} c @{} ||@{} l @{} ||@{} c @{} | @{}r @{}|@{} c @{} |@{} c @{} |@{} c @{} |}
    \hline
    L           & $r_{0}$      & $q_{1}$  & $q_{2}$ & $q_{3}$  &   $v$    & L           &  $r_{0}$        & $q_{1}$ & $q_{2}$ & $q_{3}$  &   $v$ \\ \hline\hline
    1.000001    & 1.00000025   & 0        &    0    &    0     &   0.001  & 1.68        &  0.07           &  1      &   1     &    0     &    0.1 \\ \hline
    $\infty$    &    1         & 0        &    0    &    0     &   1      & 6.37        &  1.14           &  1      &   1     &    0     &    0.9 \\ \hline
    1.000001    & 0.786        & 1        &    0    &    0     &   0.001  & 1.0067      &  3.18           & -10     &  -1     &    0     &    0.1 \\ \hline
    5.57        & 1.359        & 1        &    0    &    0     &   0.9    & 49.097      &  2.76           & -1      &   0     &    0     &    0.99 \\ \hline
    $\infty$    &    1         & 1        &    0    &    0     &   1      & 50.98       &  2.59           &  1      &   1     &   -1     &    0.99 \\ \hline
    6.56        & 0.7          & 10       &    0    &    0     &   0.9    & 1.008       &  1.3            &  1      &  -1     &   -1     &    0.1 \\  \hline
    1.01        & 0.3          & 10       &    0    &    0     &   0.1    & 48.87       &  2.77           &  1      &  -1     &   -1     &    0.99 \\  \hline
    4.8         & 1.8          & -1       &   -1    &    0     &   0.9    & 1.00008     &  1.3            &  1      &  -1     &   -1     &    0.01 \\
    \hline
  \end{tabular}
\end{center}
Table 2. Value of $L=\frac{\kappa_{\parallel}}{\kappa_{\perp}}$ for $k=0$.

\begin{center}
  \begin{tabular}{|@{} l @{} ||@{} c @{} | @{}r @{}|@{} c @{} |@{} c @{} |@{} c @{} ||@{} l @{} ||@{} c @{} | @{}r @{}|@{} c @{} |@{} c @{} |@{} c @{} |}
    \hline
    L           & $r_{0}$      & $q_{1}$  & $q_{2}$ & $q_{3}$ &   $v$     & L           &  $r_{0}$        & $q_{1}$ & $q_{2}$ & $q_{3}$  &   $v$ \\ \hline\hline
    1.000001    & 1.27         & 0        &    0    &    0    &   0.001  & 1.69         &  0.1            &  1      &   1     &    0     &    0.1 \\ \hline
    $\infty$    &    0         & 0        &    0    &    0    &   1      & 10.2         &  2.06           &  1      &   1     &    0     &    0.9 \\ \hline
    1.0000017   & 1.0000005    & 1        &    0    &    0    &   0.001  & 1.01         &  1.7            & -1      &  -1     &    0     &    0.1 \\ \hline
    9.07        & 2.29         & 1        &    0    &    0    &   0.9    & 96.5         &  7.2            & -1      &   0     &    0     &    0.99 \\ \hline
    $\infty$    &    0         & 1        &    0    &    0    &   1      & 98.8         &  7.09           &  1      &   1     &   -1     &    0.99 \\ \hline
    10.8        & 0.96         & 10       &    0    &    0    &   0.9    & 1.01         &  1.6            &  1      &  -1     &   -1     &    0.1 \\  \hline
    1.01        & 0.33         & 10       &    0    &    0    &   0.1    & 96.5         &  7.2            &  1      &  -1     &   -1     &    0.99 \\  \hline
    7.65        & 2.79         & -1       &   -1    &    0    &   0.9    & 1.01         &  1.6            &  1      &  -1     &   -1     &    0.01 \\
    \hline
  \end{tabular}
\end{center}
Table 3. Value of $L=\frac{\kappa_{\parallel}}{\kappa_{\perp}}$ for $k=-1$.\\\\

Hence, for the positive charges in relativistic regime (in the case of $k=1$), increasing value of charges and number of charges, which are corresponding to chemical potential of QGP, increase value of ratio $\frac{\kappa_{\parallel}}{\kappa_{\perp}}$.\\
The higher derivative corrections of STU model can be found in the Ref. \cite{20},
\begin{eqnarray}\label{11}
f_{k}&=&k-\frac{\mu}{r^2}+\frac{r^{2}}{R^{2}}\prod_{i}(1+\frac{q_i}{r^2})+c_{1}
\left(\frac{\mu^{2}}{96r^{6}\prod_{i}(1+\frac{q_i}{r^2})}-\frac{\prod_{i}q_{i}(q_{i}+\mu)}{9R^{2}r^{4}}\right),\nonumber\\
{\mathcal{H}}&=&\prod_{i=1}^{3} H_{i},\nonumber\\
H_{i}&=&1+\frac{q_i}{r^2}-\frac{c_{1}q_{i}(q_{i}+\mu)}{72r^{2}(r^{2}+q_{i})^{2}},
\hspace{10mm} i=1, 2, 3,
\end{eqnarray}
where $c_{1}$ is the small constant parameter corresponding to the
higher derivative terms and $a_{1}$ is $q_{i}$-dependent quantity
which parameterize the corrections to the background geometry \cite{23}.
In that case the modified horizon radius is
given by the following expression,
\begin{eqnarray}\label{12}
r_{h}&=&r_{0h}\nonumber\\
&+&\frac{c_{1}\prod_{i}(1+\frac{q_i}{r_{0h}^2})\left(\sum q_{i}^{2}-\frac{26r_{0h}^{2}}{3}\sum q_{i}+3r_{0h}^{4}\right)}{576R^{2}
\left[(\prod_{i}(1+\frac{q_i}{r_{0h}^2}))^{\frac{2}{3}}(\frac{1}{3}\sum
q_{i}-2r_{0h}^{2})-R^{2}\right]}\nonumber\\
&+&c_{1}\frac{2(\prod_{i}(1+\frac{q_{i}}{r_{0h}^2}))^{\frac{1}{3}}(\frac{13}{3}\sum
q_{i}-3r_{0h}^{2})+3R^{2}}{576
\left[(\prod_{i}(1+\frac{q_{i}}{r_{0h}^2}))^{\frac{2}{3}}(\frac{1}{3}\sum
q_{i}-2r_{0h}^{2})-R^{2}\right]}
\end{eqnarray}
where $r_{0h}$ is the horizon radius without higher derivative
corrections, it should be noted
that, in order to obtain the expression (\ref{12}) we removed $\mu$ by
using $f_{k}=0$.\\
As before, we can calculate components $G_{00}$, $G_{rr}$, $G_{kk}$ and $G_{pp}$ to obtain $\frac{\kappa_{\parallel}}{\kappa_{\perp}}$ and investigate universal behavior $\kappa_{\parallel}\geq \kappa_{\perp}$. Numerically, we find that the relation $\frac{\kappa_{\parallel}}{\kappa_{\perp}}\geq1$ is valid in presence of higher order corrections. Effect of $c_{1}$ is reduction of ratio, for example in the case of three charged black hole with $q_{1}=10^{6}$, $q_{2}=q_{3}=1$, $c_1=0.01$, and $v=0.9$ we have $\frac{\kappa_{\parallel}}{\kappa_{\perp}}=3.8$. Hence, we confirmed universal properties of the Langevin diffusion coefficients in the STU model with higher derivative terms. Already thermodynamical and statistical analysis of STU black holes given by the Ref. \cite{pour}, now it may be interesting to consider logarithmic corrected STU model \cite{LogSTU} and investigate Langevin diffusion coefficients.\\

{\bf Acknowledgments}: We would like to thanks K. Bitaghsir Fadafan for several discussions.


\begin{thebibliography}{11}
\bibitem{1}
J. M. Maldacena, "The large N limit of superconformal field theories
and supergravity", Adv. Theor. Math. Phys. 2 (1998) 231; Int. J. Theor. Phys. 38 (1999) 1113
\bibitem{2}
E. Witten, "Anti-de Sitter space and holography", Adv. Theor. Math.
Phys. 2 (1998) 253.
\bibitem{3}
S. S. Gubser, "Using string theory to study the quark-gluon plasma: progress and
perils", Nucl. Phys. A 830 (2009) 657.
\bibitem{4}
C.P. Herzog, "Energy Loss of Heavy Quarks from Asymptotically AdS Geometries", J. High Energy Phys. 0609 (2006) 032.
\bibitem{5}
S.S. Gubser, "Drag force in AdS/CFT", Phys. Rev. D, 74 (2006) 126005.
\bibitem{6}
H. Liu, K. Rajagopal, and U.A. Wiedemann, "Calculating the Jet Quenching Parameter", Phys. Rev. Lett. 97 (2006) 182301.
\bibitem{7}
B.G. Zakharov, "Radiative energy loss of high-energy quarks in finite-size nuclear matter and quark-gluon plasma", J. Exp. Theor. Phys. Lett. 65 (1997) 615.
\bibitem{8}
F.L. Lin and T. Matsuo, "Jet quenching parameter in medium with chemical potential from AdS/CFT", Phys. Lett. B 641 (2006) 45.
\bibitem{9}
S.S. Gubser, "Momentum fluctuations of heavy quarks in the gauge-string duality",  Nucl. Phys. B 790 (2008) 175.
\bibitem{10}
J. Casalderrey-Solana and D. Teaney, "Transverse Momentum Broadening of a Fast Quark in a $N=4$ Yang Mills Plasma", JHEP 0704 (2007)
039.
\bibitem{11}
J. de Boer, V.E. Hubeny, M. Rangamani and M. Shigemori, "Brownian motion in AdS/CFT",  JHEP 0907 (2009) 094.
\bibitem{12}
D.T. Son and D. Teaney, "Thermal Noise and Stochastic Strings in AdS/CFT", JHEP 0907 (2009) 021.
\bibitem{13}
D. Giataganas, H. Soltanpanahi, "Universal Properties of the Langevin Diffusion Coefficients", Phys. Rev. D 89 (2014) 026011
\bibitem{14}
D. Giataganas, H. Soltanpanahi, "Heavy Quark Diffusion in Strongly Coupled Anisotropic Plasmas", JHEP 1406 (2014) 047
\bibitem{15}
C. Hoyos-Badajoz, "Drag and jet quenching of heavy quarks in a strongly coupled N=2* plasma",  J. High Energy Phys. 0909 (2009) 068
\bibitem{16}
J. Sadeghi and B. Pourhassan, "Drag Force of Moving Quark at The N=2 Supergravity", J. High Energy Phys. 0812 (2008) 026
\bibitem{17}
J. Sadeghi, M.R. Setare, B. Pourhassan, and S. Hashmatian, "Drag Force of Moving Quark in STU Background", Eur. Phys. J. C 61
 (2009) 527
\bibitem{18}
J. Sadeghi, M.R. Setare, and B. Pourhassan, "Drag force with different charges in STU background and AdS/CFT", J. Phys. G: Nucl. Part. Phys. 36 (2009) 115005
\bibitem{19}
K. Bitaghsir Fadafan, B. Pourhassan, and J. Sadeghi, "Calculating the jet-quenching parameter in STU background", Eur. Phys. J. C 71
(2011) 1785
\bibitem{20}
B. Pourhassan, J. Sadeghi, "STU/QCD Correspondence", Canadian Journal of Physics, 91(12) (2013) 995
\bibitem{Amani}
J. Sadeghi, B. Pourhassan, A. R. Amani, "The effect of higher derivative correction on $\eta/s$ and conductivities in STU model", Int. J. Theor. Phys. 52 (2013) 42
\bibitem{S1}
H. Saadat and  B. Pourhassan, "Holographic Superfluid and STU Model", Int. J. Theor. Phys. 52 (2013) 997
\bibitem{S1}
B. Pourhassan, M.M. Bagheri-Mohagheghi, "Holographic superconductor in a deformed four-dimensional STU model", 1609.08402 [hep-th]
\bibitem{21}
V. Balasubramanian and F. Larsen, "On D-branes and black holes in four dimensions", Nucl. Phys. B 478 (1996) 199
\bibitem{22}
D.T. Son and A.O. Starinets, "Hydrodynamics of R-charged black holes", J. High Energy Phys. 0603 (2006) 052
\bibitem{23}
S. Cremonini, K. Hanaki, J.T. Liu, and P. Szepietowski, "Higher derivative effects on $\eta/s$ at finite chemical potential", Phys. Rev. D 80 (2009) 025002
\bibitem{pour}
A. Pourdarvish, B. Pourhassan, M. Tabassomi, "Statistical Analysis of STU Black Holes", Int. J. Theor. Phys. 53 (2014) 1814
\bibitem{LogSTU}
B. Pourhassan, M. Faizal, Eur. Phys. J. C 77 (2017) 96
\end{thebibliography}
\end{document}